\documentclass[journal,final,draftcls,letterpaper,oneside,twocolumn,romanappendices]{IEEEtran}
\IEEEoverridecommandlockouts
\usepackage{silence} 
\usepackage[T1]{fontenc}
\usepackage[latin9]{inputenc}
\usepackage{amsmath,amsfonts,amsthm,amssymb}
\usepackage{algorithmic,algorithm}
\usepackage{array,stackrel}
\usepackage[caption=false,font=normalsize,labelfont=sf,textfont=sf]{subfig}
\usepackage{textcomp}
\usepackage{stfloats}
\usepackage{url}
\usepackage{verbatim}
\usepackage{graphicx}
\usepackage{cite}
\usepackage{xcolor}
\usepackage{pdfcolmk}
\usepackage{refcount}
\usepackage{esint}
\usepackage{hhline}
\usepackage{bm}
\usepackage{xfrac}
\usepackage{makecell}
\usepackage{float}
\usepackage{lipsum}

\usepackage[switch]{lineno}

\usepackage{cases}
\usepackage[]{footmisc}
\def\BibTeX{{\rm B\kern-.05em{\sc i\kern-.025em b}\kern-.08em
    T\kern-.1667em\lower.7ex\hbox{E}\kern-.125emX}}

\theoremstyle{plain}
\newtheorem{thm}{\protect\theoremname}
\theoremstyle{plain}

\newtheorem{corollary}{Corollary}[thm]
\usepackage{stackengine}

\usepackage[caption=false,font=normalsize,labelfont=sf,textfont=sf]{subfig}

\makeatother

\providecommand{\lemmaname}{Lemma}
\providecommand{\theoremname}{Theorem}

\theoremstyle{definition}

\usepackage{tabularx,booktabs}
\newcolumntype{C}{>{\centering\arraybackslash}X} 
\usepackage{booktabs}
\PassOptionsToPackage{normalem}{ulem}
\usepackage{ulem}
\usepackage[unicode=true,
 bookmarks=true,bookmarksnumbered=true,bookmarksopen=true,bookmarksopenlevel=1,
 breaklinks=false,pdfborder={0 0 0},pdfborderstyle={},backref=false,colorlinks=false]
 {hyperref}
\hypersetup{pdftitle={Your Title},
 pdfauthor={Your Name},
 pdfpagelayout=OneColumn, pdfnewwindow=true, pdfstartview=XYZ, plainpages=false}
\PassOptionsToPackage{unicode}{hyperref}
\PassOptionsToPackage{naturalnames}{hyperref}
\usepackage[cmintegrals]{newtxmath}
\usepackage[caption=false,font=normalsize,labelfont=sf,textfont=sf]{subfig}

\makeatletter

\begin{document}

\title{Closed-Form Analysis of the $\alpha$-Beaulieu-Xie Shadowed Fading Channel
}

\author{Aleksey S.~Gvozdarev\href{https://orcid.org/0000-0001-9308-4386}{\includegraphics[scale=0.1]{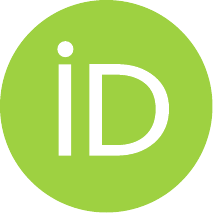}},~\IEEEmembership{Member,~IEEE,}
\thanks{\copyright 2022 IEEE.  Personal use of this material is permitted.  Permission from IEEE must be obtained for all other uses, in any current or future media, including reprinting/republishing this material for advertising or promotional purposes, creating new collective works, for resale or redistribution to servers or lists, or reuse of any copyrighted component of this work in other works.}
\thanks{The author is with the Department of Intelligent Radiophysical  Information Systems, P. G. Demidov Yaroslavl State University, 150003 Yaroslavl,
Russia (e-mail: asg.rus@gmail.com).}
\thanks{This work was supported by Russian Science Foundation under Grant 22-29-01458 (https://rscf.ru/en/project/22-29-01458/).}%
}

\maketitle

\begin{abstract}
The presented research proposes the $\alpha$-modification of the Beaulieu-Xie shadowed fading channel for wireless communications. For the assumed model the closed-form analytical description of the basic statistical characteristics is carried out (i.e., probability density function, cumulative distribution function, and their asymptotics). The derived statistical description is exemplified on the problems of the average bit error rate and ergodic capacity calculation, for which the exact analytic and high signal-to-noise ratio asymptotic expressions are derived.  The performed extensive numerical analysis  demonstrates high correspondence with the analytical work and helps to study the dependence of the channel nonlinearity effects on the bit error probability.
\end{abstract}


\section{Introduction}\label{S-1}
The evolution of the modern communication systems (like, terahertz (THz) communication, free space optical (FSO) communication, reconfigurable intelligent surfaces assisted communications, etc.) implies stringent restriction on the utilized wireless propagation channel models. Thus, their development is of primary importance for communication quality and reliability prediction. Recently, the Beaulieu-Xie model drew specific attention due to the combination of analytical tractability and good correspondence to the real-life measurements.

The classical Beaulieu-Xie model (proposed in \cite{Bea15}) presented a propagation scenario with multiple specular components, which is of primary importance for THz and  FSO communication. It was numerously used to analyze the wireless channel effective capacity \cite{Kan17}, its second-order statistics (e.g., level crossing rate and average fade duration) \cite{Olu17} and joint envelope-phase statistics \cite{Sil21}, communication error rate for DPSK/MFSK/MPSK/MQAM modulations \cite{Kan18,Kan20, Bil22}; diversity combining schemes were studied in \cite{Kau20} for the MRC receiver and in \cite{Sha22} for the switch-and-stay receiver; the physical layer security (PLS) issues were analyzed in \cite{Cha20} and in \cite{Sun22} PLS with NOMA communication and I/Q imbalance was studied.

The Beaulieu-Xie model was further elaborated in \cite{Olu20a}  to include the possible shadowing of the line-of-sight (LoS) components. The specific property of a new Beaulieu-Xie shadowed model was that it was able to handle arbitrary fading and shadowing coefficients. The novel model was analyzed from the perspectives of its error rate performance \cite{Haw22}, physical layer security \cite{Gvo22b}, and channel capacity \cite{Sil22}.

Another modern trend in wireless communications is to modify the existing channel models by accounting possible propagation nonlinearity effects, i.e., $\alpha$-variates. Starting from the pioneering work \cite{Yac07b}, where the $\alpha-\mu$ distribution was introduced, and up to the most recent ones \cite{AlH22a, AlH22b}, those models gain wide acceptance in ad hoc communications \cite{Bad20,Bha21, Li22}.

Recently, a $\alpha$-modification of the classical  Beaulieu-Xie model was presented in \cite{Che23}. Although the closed-form of the probability density function (pdf) and approximated form of the cumulative distribution function (cdf) were obtained, the complete statistical description is up to now absent. Moreover, it inherited all the limitations of the initial classical model: no shadowing, and the existing fading parameters' limitations.

To remove those drawbacks, the research presents the $\alpha$-Beaulieu-Xie shadowed fading channel model. The major contributions of this work can be summarized as follows: \textit{a)}~the novel fading model suitable for wireless communication is proposed; \textit{b)}~for the assumed channel the probability density function (pdf) and cumulative distribution functions (cdf) of the instantaneous signal-to-noise ratio (SNR) are derived; \textit{c)}~high SNR asymptotic expression of pdf and cdf are presented; \textit{d)}~the exact expressions for the average bit error rate (ABER) and ergodic capacity are derived; \textit{e)}~an asymptotic ABER and ergodic capacity are presented, and the channel diversity order is evaluated; \textit{f)}~a thorough numerical analysis verifying the correctness of the analytical work was performed, and the dependence of the system performance quality on the channel parameters is studied.

The rest of the paper is organized as follows: in Section~\ref{S-2} a brief description of the initial Beaulieu-Xie shadowed model is given; Section~\ref{S-3} presents the statistical analysis of the proposed $\alpha$-Beaulieu-Xie shadowed model; Section~\ref{S-4} derives the exact expressions of the average bit error probability, ergodic capacity and their asymptotics; Section~\ref{S-5} presents a thorough numerical analysis of the derived expressions depending on various channel and system parameters' values; and the conclusions are drawn in Section~\ref{S-6}.

\section{Beaulieu-Xie shadowed model}\label{S-2}
The Beaulieu-Xie shadowed \cite{Olu20a} is a four-parametric fading channel model with the probability density function of the signals' envelope, given by \cite{Olu20a}:
\begin{IEEEeqnarray}{rCl}\label{BXs:pdf}
f_{\bar{R}}(\bar{r})&=&\frac{2 \bar{r}^{2m_X-1}}{\Gamma (m_X)}\left(\frac{m_Y\Omega_X}{m_Y\Omega_X+m_X\Omega_Y}\right)^{m_Y} \left(\frac{m_X}{\Omega_X}\right)^{m_X}   \times    \IEEEnonumber \\
&&\hspace{-20pt}   \times   \mbox{}_1F_1\left(m_Y, m_X, \frac{m_X^2\Omega_Y \bar{r}^2}{\Omega_X(m_Y\Omega_X+m_X\Omega_Y)}\right)e^{-\frac{m_X}{\Omega_X}\bar{r}^2},
\end{IEEEeqnarray}
where $\Gamma (   \cdot )$ is the Euler gamma-function \cite{Gra96}, and $\mbox{}_1F_1(   \cdot )$ is the Kummer's confluent hypergeometric function \cite{Gra96}. The parameters $m_X, m_Y$ define the overall (LoS and NLoS) signal fading and LoS components' shadowing, respectively; and $\Omega_X, \Omega_Y$ quantify NLoS and LoS components' power, respectively. It generalizes a wide range of classical fading models \cite{Olu20a}, including Rayleigh, Rice, Rician shadowed, one-sided Gaussian, Nakagami-m, $\kappa-\mu$, $\kappa-\mu$ shadowed, $\eta-\mu$, etc. Importantly, the parameters of the models can take arbitrary positive values, in contrast, for example, to the widely used Nakagami-related models, where the shadowing and fading parameters are lower-bounded with $1/2$.

\section{$\alpha$-Beaulieu-Xie shadowed modification}\label{S-3}

To derive the $\alpha$-modification of the channel model \eqref{BXs:pdf}, one assumes the classical approach \cite{Yac07b} within which it is supposed that the signal propagating in a nonhomogeneous environment experiences nonlinear envelope distortions.

To this extent, applying no restrictions (with the exception of positiveness) neither on channel parameters $m_X, m_Y, \Omega_X, \Omega_Y$ nor on the propagation medium nonlinearity, $\alpha$ the following statements concerning the basic statistical description are derived.

\begin{thm} The probability density function $f_{\gamma }(\gamma )$ and the cumulative distribution function $F_{\gamma }(\gamma )$ of the instantaneous signal-to-noise ratio $\gamma$ for the $\alpha$-Beaulieu-Xie shadowed fading channel model with arbitrary positive channel parameters (i.e., $m_X, m_Y, \Omega_X, \Omega_Y, \alpha \geq 0$) can be expressed as
\begin{IEEEeqnarray}{rCl}
f_\gamma (\gamma )&=&\frac{\alpha \left(1-\bar{\beta }\right)^{m_Y} }{2\mathrm{C}_{\alpha }^{m_X}\Gamma (m_X)\bar{\gamma }}\left(\frac{\gamma }{\bar{\gamma }}\right)^{\frac{\alpha m_X}{2}-1} e^{-\mathrm{C}_{\alpha }^{-1}\left(\frac{\gamma }{\bar{\gamma }}\right)^{\frac{\alpha }{2}}}   \times    \IEEEnonumber \\
&&   \times   \mbox{}_1F_1\left(m_Y, m_X, \frac{\bar{\beta }}{\mathrm{C}_{\alpha }}\left(\frac{\gamma }{\bar{\gamma }}\right)^{\frac{\alpha }{2}}\right),\label{thm:1.1}\\
F_\gamma (\gamma )&=&\frac{\left(1-\bar{\beta }\right)^{m_Y}}{\mathrm{C}_{\alpha }^{m_X}\Gamma (m_X+1)}\left(\frac{\gamma }{\bar{\gamma }}\right)^{\frac{\alpha m_X}{2}} e^{-\mathrm{C}_{\alpha }^{-1}\left(\frac{\gamma }{\bar{\gamma }}\right)^{\frac{\alpha }{2}}}    \times    \IEEEnonumber \\
&&   \times   \; \Phi_2\left(1, m_Y; m_X; \frac{1}{\mathrm{C}_{\alpha }}\left(\frac{\gamma }{\bar{\gamma }}\right)^{\frac{\alpha }{2}}, \frac{\bar{\beta }}{\mathrm{C}_{\alpha }}\left(\frac{\gamma }{\bar{\gamma }}\right)^{\frac{\alpha }{2}}\right),
\label{thm:1.2}
\end{IEEEeqnarray}
where $\mathrm{C}_{\alpha }=\left[\frac{\Gamma (m_X)}{\Gamma \left(m_X+\frac{2}{\alpha }\right)\mbox{}_2F_1\left(m_Y,-\frac{2}{\alpha }; m_X; -\frac{m_X\Omega_Y}{m_Y\Omega_X}\right)}\right]^{\frac{\alpha }{2}}$, $\bar{\gamma }$ is the average signal-to-noise ratio, $\bar{\beta }=\left(\frac{m_X\Omega_Y}{m_Y\Omega_X+m_X\Omega_Y}\right)$, $\mbox{}_2F_1(   \cdot )$ is the Gauss hypergeometric function \cite{Gra96}, $\Phi_2(   \cdot )$ is the bivariate confluent Appell function \cite{Gra96}.
\end{thm}

\begin{IEEEproof}
To prove Theorem 1, first let us note that \eqref{BXs:pdf} defines the non-normalized envelope $\bar{r}$ pdf. Thus, complying with the procedure presented in \cite{Yac07b}, one can assume the following change of variable: $r^2=\bar{r}^2\mathbb{E}\{R^2\}/\mathbb{E}\{\bar{R}^2\}$, where $\mathbb{E}\{   \cdot \}$ is the averaging operator. The second moment of $\bar{R}$ (i.e., $\mathbb{E}\{\bar{R}^2\}$) can be derived by utilizing the fact that
\begin{equation}\label{pf:1.1}
\mathbb{E}\{\bar{R}^k\}=\frac{\mbox{}_2F_1\left(m_Y,-\frac{k}{2}; m_X; -\frac{m_X\Omega_Y}{m_Y\Omega_X}\right)}{\Gamma (m_X)\Gamma \left(m_X+\frac{k}{2}\right)^{-1} } \left(\frac{m_X}{\Omega_X}\right)^{-\frac{k}{2}},
\end{equation}
which leads to $\mathbb{E}\{\bar{R}^2\}=\Omega_X+\Omega_Y$. Next, one derives the pdf for the power of the normalized envelope $Z=R^2$ via random variable transformation technique, i.e., $f_Z(z)=\frac{1}{2}z^{-\frac{1}{2}}f_R(z^{\frac{1}{2}})$, and converting it in to the $\alpha$-root envelope, i.e.,  $f_{R^{\alpha }}(r)=\frac{\alpha }{2}r^{\alpha-1}f_Z(r^{\alpha })$  with the mean-square (i.e., $\mathbb{E}\{\left(R^{\alpha }\right)^2\}=\mho_{\alpha }$)
\begin{equation}\label{pf:1.2}
    \mho_{\alpha }=\left[\frac{\Omega_X\mathbb{E}\{R^2\}}{m_X\mathrm{C}_{\alpha }(\Omega_X+\Omega_Y)}\right]^{\frac{2}{\alpha }}.
\end{equation}

Lastly, the envelope of the instantaneous SNR $\gamma$ is obtained via transformation $f_{\gamma }(\gamma )=\frac{1}{2}\sqrt{\frac{\mho_{\alpha }}{\gamma \bar{\gamma }}}f_{R^{\alpha }}\left(\sqrt{\frac{\mho_{\alpha }\gamma }{\bar{\gamma }}}\right)$, which finalizes the proof of \eqref{thm:1.1}.

To prove \eqref{thm:1.2}, one starts with the definition of the cumulative distribution function, and perform the change of variable $x^{\frac{\alpha }{2}}/\gamma^{\frac{\alpha }{2}}=t$, which yields:
\begin{IEEEeqnarray}{rCl}\label{pf:1.3}
F_\gamma (\gamma )&=&  \frac{\left(1-\bar{\beta }\right)^{m_Y}}{\mathrm{C}_{\alpha }^{m_X}\Gamma (m_X)}\left(\frac{\gamma }{\bar{\gamma }}\right)^{\frac{\alpha }{2}}   \times    \IEEEnonumber \\
&&\hspace{-45pt}   \times   \int_0^1
\frac{
e^{-\mathrm{C}_{\alpha }^{-1}\left(\frac{\gamma }{\bar{\gamma }}\right)^{\frac{\alpha }{2}}t}
}{t^{1-m_X}}  \mbox{}_1F_1\left(m_Y;m_X; \frac{\bar{\beta }}{\mathrm{C}_{\alpha }}\left(\frac{\gamma }{\bar{\gamma }}\right)^{\frac{\alpha }{2}}t\right){\rm{d}}t.
\end{IEEEeqnarray}

After reorganizing the terms, the application of the integral representation for the confluent Appell function (see expression $(3.5)$ in \cite{Bry12}) finalizes the proof of Theorem 1.
\end{IEEEproof}

It can be easily verified that substituting $\alpha=2$ in \eqref{thm:1.1}-\eqref{thm:1.2} and performing simplifications yields the corresponding expressions for the initial Beaulieu-Xie shadowed model. The obtained results in the high-SNR region can be simplified as follows.

\begin{corollary}
The asymptotic expressions of pdf and cdf ($\bar{\gamma }\to \infty$) for the $\alpha$-Beaulieu-Xie shadowed fading channel model is given by\footnote{The proof is obtained by observing that $\displaystyle \lim_{x\to \infty} \mbox{}_1{}F_1(a;b;x^{-1})=1$ and  $\displaystyle \lim_{x\to \infty} \Phi_2(a,b;c;x^{-1},x^{-1})=1$}
\begin{IEEEeqnarray}{rCl}
f_\gamma (\gamma )&\approx&\frac{\alpha \left(1-\bar{\beta }\right)^{m_Y} }{2\mathrm{C}_{\alpha }^{m_X}\Gamma (m_X)\bar{\gamma }}\left(\frac{\gamma }{\bar{\gamma }}\right)^{\frac{\alpha m_X}{2}-1} e^{-\mathrm{C}_{\alpha }^{-1}\left(\frac{\gamma }{\bar{\gamma }}\right)^{\frac{\alpha }{2}}},\label{cor:1.1}\\
F_\gamma (\gamma )&\approx&\frac{\left(1-\bar{\beta }\right)^{m_Y}}{\mathrm{C}_{\alpha }^{m_X}\Gamma (m_X+1)}\left(\frac{\gamma }{\bar{\gamma }}\right)^{\frac{\alpha m_X}{2}}.
\label{cor:1.2}
\end{IEEEeqnarray}
\end{corollary}

The derived statistical description of the $\alpha$-Beaulieu-Xie shadowed fading channel model enables the evaluation of the system performance metrics.

\section{Performance analysis for the $\alpha$-Beaulieu-Xie shadowed model}\label{S-4}

To illustrate the performance of the wireless communication system functioning in the presence of $\alpha$-Beaulieu-Xie shadowed fading channel, one assumes the average bit error rate (ABER) and ergodic capacity as quantifying metrics.
\subsection{ABER performance in the presence of the $\alpha$-BX shadowed model}\label{S-4.1}

For a wide range of commonly used modulation schemes \cite{AlH22b} (i.e., for example, BPSK, GMSK (for high $\bar{\gamma }$), M-PSK, M-FSK, square M-QAM, and M-DPSK), ABER can be efficiently approximated as the weighted sum of the averaged Gauss Q-functions $Q(   \cdot )$, i.e.
\begin{IEEEeqnarray}{rCl} \label{eq:ABER}
&&\mathrm{\overline{P}_{er}}=\delta_{1}\sum_{j=1}^{\delta_{3}}\int_{0}^{\infty}Q(\sqrt{2\delta_{2,j}\gamma })f_{\gamma }(\gamma ){\rm d}\gamma,
\end{IEEEeqnarray}
where the set of coefficients $\left\{ \delta_{1},\delta_{2,j},\delta_{3}\right\} $ are explicitly defined for the specific modulation (see, for instance, \cite{Lu99}). For example, for M-QAM $\left\{ \frac{4\left(1-\sfrac{1}{\sqrt{M}}\right)}{\log_2 M}, \frac{3(2j-1)^2}{2(M-1)}, \frac{\sqrt{M}}{2}\right\}$.

\begin{thm}
The ABER for the $\alpha$-Beaulieu-Xie shadowed fading channel model with arbitrary positive channel parameters can be expressed as
\begin{IEEEeqnarray}{rCl}\label{thm:2}
\mathrm{\overline{P}_{er}}&=&\frac{\delta_{1}\sqrt{\pi }\left(1-\bar{\beta }\right)^{m_Y}}{q^{m_X-\frac{p+q}{2}}\left(2\pi \right)^{\frac{p+q}{2}}}\sum_{j=1}^{\delta_{3}}\sum_{k=0}^{\infty}\frac{(m_Y)_{k}}{\Gamma (m_X+k)}\frac{\left(q\bar{\beta }\right)^k}{k!} \IEEEnonumber   \times    \\
&&    \times     G_{p+1,q+1}^{\,q,p+1}\!\left(\left.
\frac{\left(p \delta_{2,j}^{-1}\right)^p}{\left(q \mathrm{C}_{\alpha } \bar{\gamma }^{\frac{\alpha }{2}}\right)^q}\right|\,{\begin{matrix}\Delta \left(p,\frac{1}{2}\right),1\\ \Delta (q,m_X+k),0\end{matrix}}\;
\right),
\end{IEEEeqnarray}
where $G(   \cdot )$ is the Meijer G-function \cite{Gra96}, $\Delta (a,b)=\{\frac{b}{a},\ldots,\frac{b+a-1}{a}\}$, $p$ and $q$ are such that $\alpha/2 = p/q$ and $\mathrm{gcd}(p, q) = 1$.
\end{thm}
\begin{IEEEproof}
To prove Theorem 2, first, let us note that
\begin{equation}  \label{pf:2.1}  \int_{0}^{\infty}Q(\sqrt{2\delta_{2,j}\gamma })f_{\gamma }(\gamma ){\rm d}\gamma=\frac{1}{2}\sqrt{\frac{\delta_{2,j}}{\pi }}\int_{0}^{\infty}\frac{e^{-\delta_{2,j}\gamma }}{\sqrt{\gamma }}F_{\gamma }(\gamma ){\rm d}\gamma.
\end{equation}

Using the expansion of the bivariate confluent Appell function $\Phi_2(   \cdot )$ in terms of the univariate $\mbox{}_1F_1(   \cdot )$ (see expression $(4.19)$ in \cite{Bry12}) and the relation  $(8.356.3)$ in \cite{Gra96}), cdf $F_{\gamma }(\gamma )$ can be rewritten in the following simple form:
\begin{equation}   \label{pf:2.2}
F_{\gamma }(\gamma )=(1-\bar{\beta })^{m_Y}\sum_{k=0}^{\infty}\frac{(m_Y)_{k}\Tilde{\gamma }\left(m_X+k,\frac{1}{\mathrm{C}_{\alpha }}\!\left(\frac{\gamma }{\bar{\gamma }}\right)^{\!\!\frac{\alpha }{2}}\right)}{\Gamma (m_X+k)}\frac{\bar{\beta }^k}{k!},
\end{equation}
where $\Tilde{\gamma }(a,z)$ is the lower incomplete gamma-function \cite{Gra96}. Combining \eqref{pf:2.1}-\eqref{pf:2.2}, and treating the obtained integral as the Laplace transform, by virtue of $(3.10.1.8)$ from \cite{Gra96}, one obtains \eqref{thm:2}.
\end{IEEEproof}

Applying the results of Corollary 1, the high-SNR behavior of \eqref{thm:2} can be obtained as follows.

\begin{corollary}
    The asymptotic expression of the ABER \eqref{thm:2} for the high-SNR region is given by
\begin{IEEEeqnarray}{rCl}\label{cor:2.1}
\mathrm{\overline{P}_{er}}\big|_{\bar{\gamma }\to\infty}&\approx &
\frac{\delta_{1}\left(1-\bar{\beta }\right)^{m_Y}\Gamma \left(\frac{\alpha m_X+1}{2}\right)}{2\sqrt{\pi }\mathrm{C}_{\alpha }^{m_X}\Gamma (m_X+1)\bar{\gamma }^{\frac{\alpha m_X}{2}}}\sum_{j=1}^{\delta_{3}}\delta_{2,j}^{-\frac{\alpha m_X}{2}}.
\end{IEEEeqnarray}
\end{corollary}

The asymptotic ABER can be expressed in the following form $\mathrm{\overline{P}_{er}}\big|_{\bar{\gamma }\to\infty}=\mathcal{G}_c\bar{\gamma }^{-\mathcal{G}_d}$, where $\mathcal{G}_d$, $\mathcal{G}_c$ are the diversity order and the coding gain.

\begin{corollary}
    The diversity order for the $\alpha$-Beaulieu-Xie shadowed fading channel model linearly scales with the power of the envelope, and does not depend on the LoS shadowing, i.e., $\mathcal{G}_d=\frac{\alpha }{2} m_X$.
\end{corollary}

\begin{figure}[!t]
\centerline{\includegraphics[width=\columnwidth]{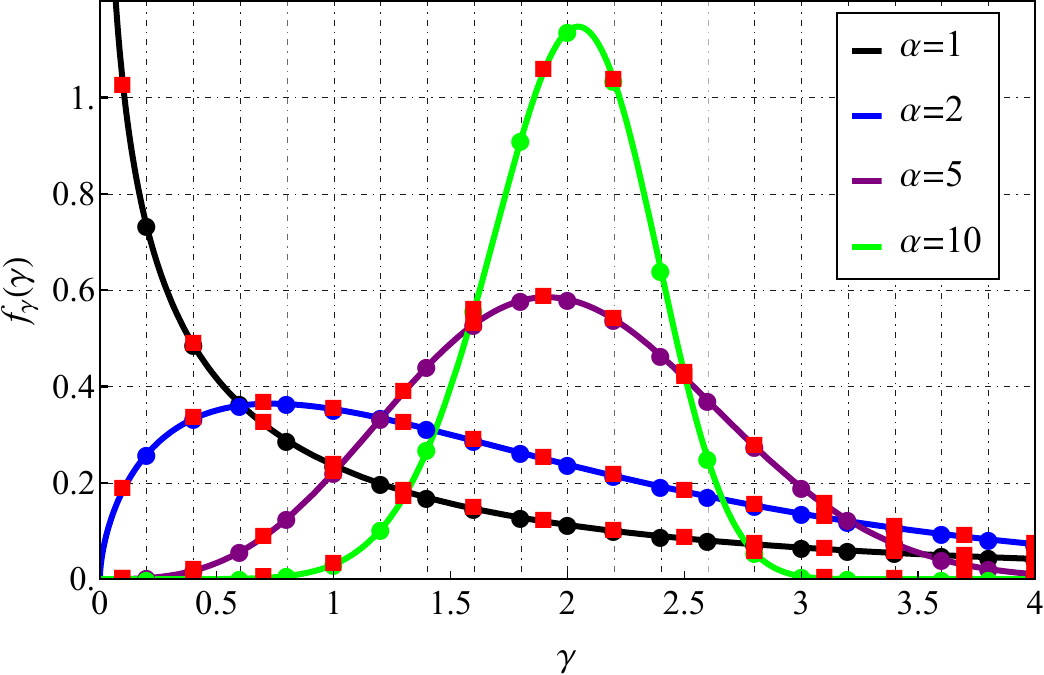}}
\caption{Instantaneous SNR pdf for $\bar{\gamma }=3$~dB, $\Omega_X=\Omega_Y=2$~dB, $m_X=1.6$, $m_Y=1.5$, and various $\alpha$: solid lines - the proposed analytical expression \eqref{thm:1.1}, circle markers - numeric simulation, and red square markers - the derived asymptotic pdf \eqref{cor:1.1}.}
\label{fig1}
\end{figure}

\subsection{Capacity analysis of the $\alpha$-BX shadowed model}\label{S-4.2}

The classical expression for the ergodic capacity of the wireless communication link subjected to multipath fading is defined as the expectation of the Shannon's capacity over the instantaneous SNR realizations, i.e.,
\begin{IEEEeqnarray}{rCl} \label{eq:Cap}
&&\mathrm{\overline{C}}=\int_{0}^{\infty}\log_2(1+\gamma )f_{\gamma }(\gamma ){\rm d}\gamma.
\end{IEEEeqnarray}

Thus, the  derived in Section~\ref{S-3} statistical representation helps to find the exact expression for $\mathrm{\overline{C}}$.

\begin{thm}
The ergodic capacity for the $\alpha$-Beaulieu-Xie shadowed fading channel model with arbitrary positive channel parameters can be expressed as
\begin{IEEEeqnarray}{rCl}\label{thm:3}
\mathrm{\overline{C}}&=&\frac{q^{m_X-\frac{1}{2}}\left(1-\bar{\beta }\right)^{m_Y}}{\left(2\pi \right)^{\frac{q-3}{2}+p}\ln2}\sum_{k=0}^{\infty}\frac{(m_Y)_{k}}{\Gamma (m_X+k)}\frac{\left(q\bar{\beta }\right)^k}{k!} \IEEEnonumber   \times    \\
&&\hspace{-30pt}    \times     G_{p+1,q+p+1}^{\,q+p+1,p}\!\left(\left.\left(\frac{1}{ q \mathrm{C}_{\alpha } \bar{\gamma }^{\frac{\alpha }{2}}}\right)^{q}\right|\,{\begin{matrix}\Delta \left(p,0\right),1\\ \Delta \left(p,0\right),\Delta (q,m_X+k),0\end{matrix}}\;
\right),
\end{IEEEeqnarray}
with $G(   \cdot )$, $\Delta (a,b)$, $p$ and $q$ defined as in Theorem 1.
\end{thm}
\begin{IEEEproof}
To prove Theorem 3, first, let us note that
\begin{equation}  \label{pf:3.1}  \int_{0}^{\infty}\log_2(1+\gamma )f_{\gamma }(\gamma ){\rm d}\gamma = \frac{1}{\ln2}\int_{0}^{\infty}\frac{\bar{F}_{\gamma }(\gamma )}{1+\gamma }{\rm d}\gamma,
\end{equation}
where $\Tilde{F}_{\gamma }(\gamma )$ denotes the complementary CDF (CCDF), i.e., $\Tilde{F}_{\gamma }(\gamma )=1-F_{\gamma }(\gamma )$.
In \eqref{pf:2.2}, using the relation between the lower $\Tilde{\gamma }(a,z)$ and the upper $\Gamma (a,z)$ incomplete gamma-functions (see $(8.2.3)$ and the result $(5.2.11)$ from \cite{Pru86}), CCDF can be rewritten in the following form:
\begin{equation}   \label{pf:3.2}
\Tilde{F}_{\gamma }(\gamma )=(1-\bar{\beta })^{m_Y}\sum_{k=0}^{\infty}\frac{(m_Y)_{k}\Gamma \left(m_X+k,\frac{1}{\mathrm{C}_{\alpha }}\!\left(\frac{\gamma }{\bar{\gamma }}\right)^{\!\!\frac{\alpha }{2}}\right)}{\Gamma (m_X+k)}\frac{\bar{\beta }^k}{k!}.
\end{equation}

Transformation of the integrands into the Meijer G-functions, i.e., $\Gamma (a,z)=G_{1,2}^{\,2,0}\!\left( z \Big|\,{\begin{matrix}1\\ 0, a\end{matrix}}\;
\right)$ (see $8.4.16.2$ in \cite{Pru90}), and $(1+z)^{-1}=G_{1,1}^{\,1,1}\!\left( z \Big|\,{\begin{matrix}0\\ 0\end{matrix}}\;
\right)$ (see $(8.4.2.5)$ in \cite{Pru90}) makes it possible to apply $(2.24.2.4)$ from \cite{Pru90}. Finally, noting that some parameters of the resultant Meijer G-function cancel out, the expression \eqref{thm:3} is obtained.
\end{IEEEproof}




\begin{thm}
The high-SNR approximation of the ergodic capacity for the $\alpha$-Beaulieu-Xie shadowed fading channel model with arbitrary positive channel parameters can be expressed as
\begin{IEEEeqnarray}{rCl}\label{thm:4}
\mathrm{\overline{C}}\big|_{\bar{\gamma }\to\infty}&\approx&\frac{2}{\alpha \ln2}\left(\ln\left( \mathrm{C}_{\alpha } \bar{\gamma }^{\frac{\alpha }{2}}\right) + \psi \left(m_X\right) \IEEEnonumber+ \right. \\
&& \left.+(1-\bar{\beta })^{m_Y}\mbox{}_2F_1^{(1,0,0,0)}\left(m_X,m_Y;m_X;\bar{\beta }\right)\right),
\end{IEEEeqnarray}
where $\psi (   \cdot )$ is the digamma-function \cite{Gra96}, and $\mbox{}_2F^{(1,0,0,0)}_1(   \cdot )$ is the derivative of  $\mbox{}_2F_1(   \cdot )$ with the respect to the first parameter.
\end{thm}
\begin{IEEEproof}
To prove Theorem 4, first, note that for the high-SNR regime $\mathrm{\overline{C}}\left.\!\!\right|_{\bar{\gamma }\to\infty}\approx\int_{0}^{\infty}\log_2(\gamma )f_{\gamma }(\gamma ){\rm d}\gamma$. Next, one expresses the logarithmic multiplier in the form of a derivative, i.e., $\log_2(\gamma )=\frac{1}{\ln2}\frac{\partial \gamma^s}{\partial s}\left.\!\!\right|_{s=0}$, substitutes \eqref{thm:1.1}, and performs the change of variable $\gamma^{\frac{\alpha }{2}}=t$. The obtained integral can be treated as a Laplace transform. Thus, applying expression $(3.35.1.2)$ from \cite{Pru92}, performing differentiation, and setting $s=0$ finalizes the proof.
\end{IEEEproof}


One can note that all the special functions in Theorems~1-3 and Corollaries  are readily accessible in all modern computer software for analytic and numeric calculations. This also includes the derivative of the hypergeometric function in \eqref{thm:4}, since in most modern software it can be easily treated both numerically and analytically. Moreover, extensive numeric analysis demonstrated that in almost all practical scenarios, the infinite series in \eqref{thm:2} can be efficiently (with the 5 digit accuracy) truncated by first $3-7$ terms.

To the best of the author's knowledge, the $\alpha$-Beaulieu-Xie shadowed fading model has not been reported in technical literature yet, and the derived results of its statistical description \eqref{thm:1.1} - \eqref{cor:1.2}, \eqref{thm:2}, \eqref{cor:2.1}, \eqref{thm:3}, \eqref{thm:4} are novel.

\begin{figure}[!t]
\centerline{\includegraphics[width=\columnwidth]{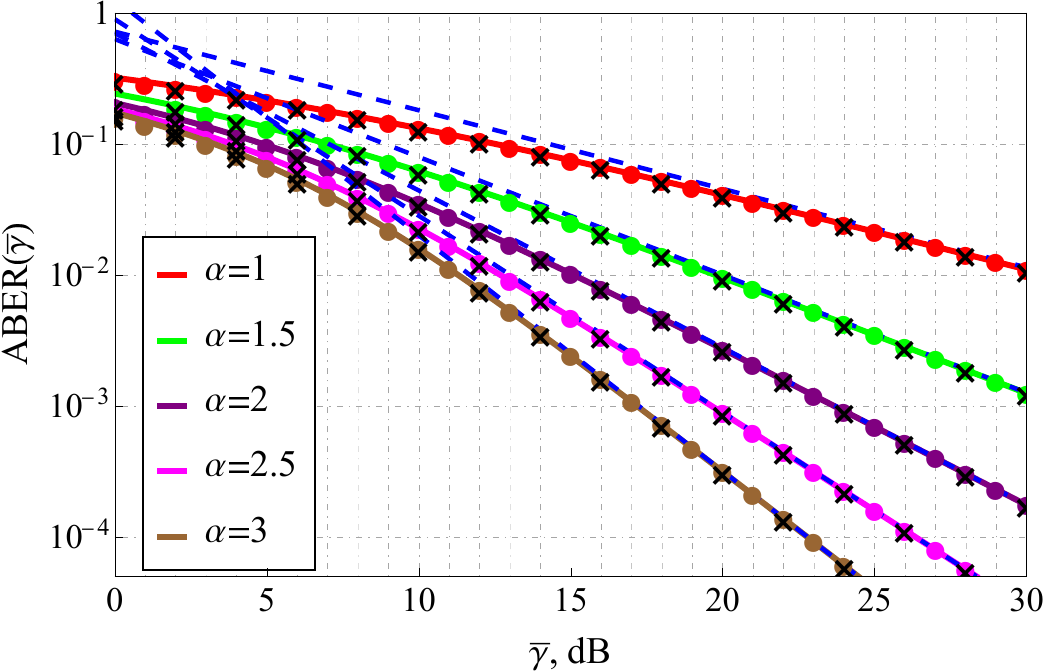}}
\caption{QAM-16 ABER for various $\bar{\gamma }$ and $m_x=m_Y=1.2$,  $\Omega_X=\Omega_Y=1$~dB: solid lines - proposed analytical expression \eqref{thm:2}, dashed lines - asymptotic expression \eqref{cor:2.1}, colored circle markers - numeric integration in \eqref{eq:ABER}, and black crossed markers - numeric simulation.}
\label{fig2}
\end{figure}
\begin{figure}[!t]
\centerline{\includegraphics[width=\columnwidth]{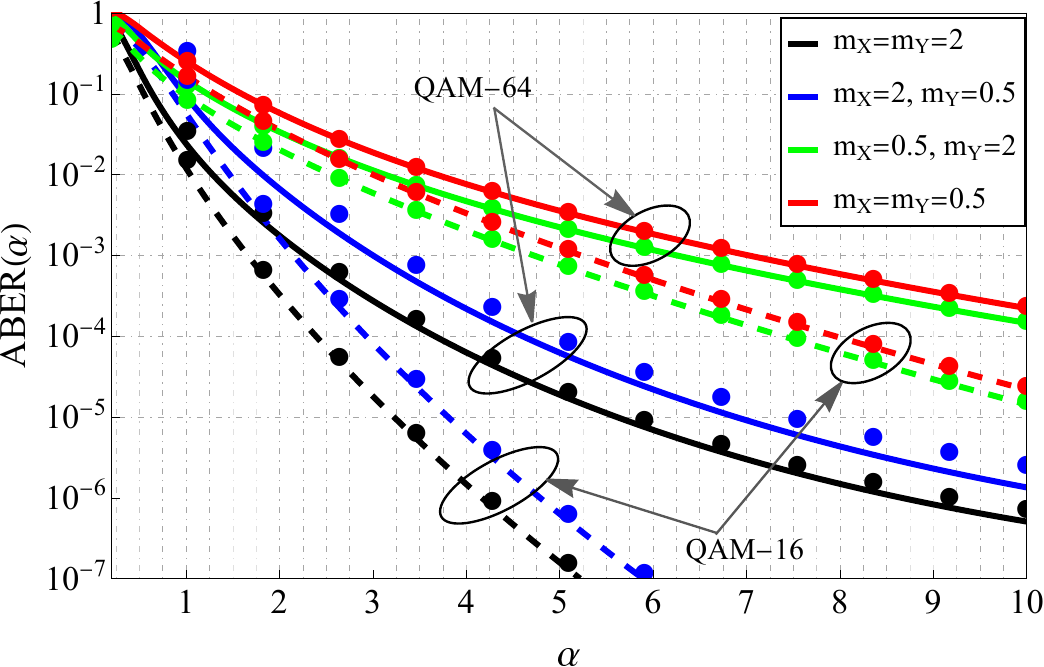}}
\caption{ABER for various $\alpha$ with $\bar{\gamma }=20$~dB, $\Omega_X=-3$~dB, $\Omega_Y=3$~dB: lines (solid and dashed) - the proposed analytical expression \eqref{thm:2}, colored circle markers - the derived asymptotic expression \eqref{cor:2.1}.}
\label{fig3}
\end{figure}

\begin{figure}[!t]
\centerline{\includegraphics[width=\columnwidth]{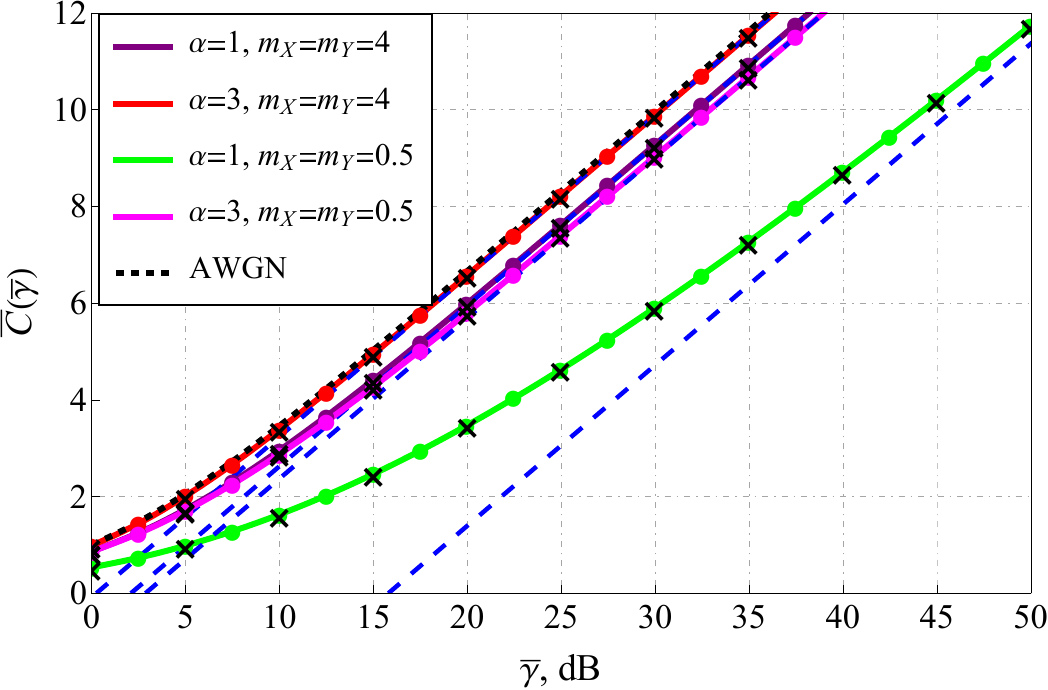}}
\caption{Capacity for various $\bar{\gamma }$ and fading conditions with  $\Omega_X=\Omega_Y=1$~dB: solid lines - proposed analytical expression \eqref{thm:3}, dashed blue lines - asymptotic expression \eqref{thm:4}, colored circle markers - numeric integration in \eqref{eq:Cap}, black crossed markers - numeric simulation, and black dashed line - AWGN channel.}
\label{fig4}
\end{figure}

\section{Simulation and results}\label{S-5}
To verify the correctness of the derived results and study the performance of the wireless communication system subjected to $\alpha$-Beaulieu-Xie shadowed fading, numerical simulation was performed.

The presented expression for instantaneous SNR pdf \eqref{thm:1.1} and its asymptotics \eqref{cor:1.1} were compared with the simulation results (see Fig.~\ref{fig1}), and an excellent agreement between the two was observed. It can be noted that for the low values of $\alpha$, the mode of the distribution is around $0$, whereas the increase of $\alpha$ leads to the pronounced unimodal behavior.

The derived ABER expressions \eqref{thm:2}, \eqref{cor:2.1} were exemplified on an M-QAM modulation (see Fig.~\ref{fig2}-\ref{fig3}).  Fig.~\ref{fig2} combines the results for QAM-16 (with moderate fading and shadowing): the derived exact solution \eqref{thm:2} (truncated to $3$ terms), its asymptotics \eqref{cor:2.1}, numeric integration in \eqref{eq:ABER}, and Monte-Carlo simulation (with $10^7$ trials), which excellently correspond each other.
As it was mentioned earlier, the increase of the nonlinearity coefficient $\alpha$ leads to the decrease of the instantaneous SNR fluctuations (i.e., the existence of the pronounced extremum), which leads to the gain in ABER. In the high-SNR region, this fact can be proved by the results obtained in Corollary~2.2.

Numerical simulation demonstrated that for the integer values of $\alpha$, the derived expression \eqref{thm:2} can deliver a sufficient computational speed-up. For example, setting the accuracy of ABER evaluation on the level of $1\%$, for QAM-$16$ modulation with both integer ($m_X=m_Y=1$, $\Omega_X=\Omega_Y=0$~dB) and non-integer ($m_X=m_Y=0.5$, $\Omega_X=\Omega_Y=1$~dB) sets of fading parameters, the proposed solution was on average $1.9-2.3$ times faster than the direct numerical integration for $-30$~dB$<\!\!\bar{\gamma }\!<\!10$~dB and $2.4-3$ times faster for $10$~dB$<\bar{\gamma }<50$~dB.

To study the impact of the nonlinearity coefficient, all the possible channel fading/shadowing conditions were assumed, e.g., strong/weak overall fading with LoS heavy/weak shadowing (see Fig.~\ref{fig3}). It can be seen that the channel model exhibits greater sensitivity to the overall fading (i.e., $m_X$), rather than to the LoS components' shadowing ($m_Y$), even in the case of strong LoS (large $\Omega_Y$). Moreover, the coding gain $\mathcal{G}_c$ (deduced from \eqref{cor:2.1}) delivers smaller improvement for heavy fading, but helps to reduce ABER with constellation size reduction. This is due to the fact that $\mathcal{G}_c$ depends not only on the channel parameters, but on the modulation dimension. It is also evident that the derived asymptotics \eqref{cor:2.1} (depicted with points) only upper-bounds the exact expression \eqref{thm:2}.

The performed capacity analysis (see Fig.~\ref{fig4}) demonstrates that $\alpha$ has greater impact for strong fading and high shadowing. Moreover, its increase helps to combat impairments induced by fading (e.g., purple and magenta lines in Fig.~\ref{fig4} are very close).  It is worth noting that in the case of fine propagation conditions with high $\alpha$ (see red line in Fig.~\ref{fig4}), the ergodic capacity almost reaches its value for the AWGN channel (black dashed line).

\section{Conclusion}\label{S-6}

The research proposes the $\alpha$-modification of the Beaulieu-Xie shadowed fading channel. For the proposed model, the general statistical description (i.e., pdf, cdf) of the instantaneous SNR and its asymptotics are derived and analyzed. The obtained channel model is exemplified on the problems of average bit probability and ergodic capacity calculation. By means of the presented results, an exact and asymptotic expressions of ABER and ergodic capacity are derived. The performed numerical analysis verified the correctness of the analytical work and helped to study the impact of the nonlinearity and the channel parameters on ABER.

\bibliographystyle{IEEEtran}
\bibliography{IEEEabrv,WCL}

\end{document}